\documentclass{LT23auth}
\usepackage{graphicx}

\begin{document}

\begin{frontmatter}

\title{On Berry phase in Bloch states}

\author[address1]{Jun Goryo\thanksref{thank1}},
\author[address1]{Mahito Kohmoto}

\address[address1]{Institute for Solid State Physics, University 
of Tokyo, 5-1-5 Kashiwanoha,
 Kashiwa, Chiba 277-8581, Japan}

\thanks[thank1]{Corresponding author. Address after Sept. 2002: Max Planck 
Institute for the Physics of Complex Systems, N\"{o}thnizer
 Street 38, 01187 Dresden, Germany  
 E-mail: jungoryo@mpipks-dresden.mpg.de, jungoryo@hotmail.com}

\begin{abstract}
We comment on the relation between Berry phase and 
quantized Hall conductivities for charge and spin currents 
in some Bloch states, such as Bloch electrons in the 
presence of electromagnetic fields and quasiparticles 
in the vortex states of superfluid $^3$He. One can find out 
that the arguments presented here are closely related to the 
spontaneous polarization in crystalline dielectrics 
and the adiabatic pumping.

\end{abstract}

\begin{keyword}
Berry phase, Bloch states, (spin) quantum Hall effect, Chern 
number, spontaneous polarization in crystalline dielectrics, 
adiabatic pumping 
 
\end{keyword}
\end{frontmatter}
\section{Introduction}
Berry revealed that a geometrical phase 
(Berry phase) arises from the adiabatic 
process of a quantum mechanical system 
around a closed loop in the parameter space. 
In spite of the fact that it is a phase of the wave function,
it could be related to physical effects and, in some cases,
has a connection with topological numbers.  Recently, 
the Berry phase is argued in the context of crystalline dielectrics 
by King-Smith and Vanderbilt, and Resta.  
The adiabatic change of the Kohm-Sham potential was considered.  
It was shown that the polarization change occurs spontaneously (i.e. 
the electric field is held to be zero) and it is written by 
the Berry phase. In this paper, we argue the system 
in the presence of electromagnetic 
field. 
It is renowned that the quantized Hall effect (QHE) occurs in 
such systems in two dimensions (2D) and also 3D. 
We define the electric polarization 
and calculate the polarization change under the 
adiabatic change of the vector potential. Then, we can find out 
the similarity to the spontaneous polarization and also the adiabatic 
pumping, which is originally argued by Thouless and 
discussed actively in the mesoscopic systems at present. 
We also show that a parallel discussion 
can be made for the vortex states of superfluid $^3$He in 2D.  
The detailed discussions and references are written in 
Refs.\cite{Goryo-Kohmoto}.

\section{Polarization in Bloch electrons in the presence of 
electromagnetic fields}
First, we argue 2D system. 
Consider the electrons under a periodic potential $U({\bf r})
=U({\bf r} + {\bf e}_x a)=U({\bf r} + {\bf e}_y a)$, here  
we consider the square lattice for simplicity.  
We introduce uniform electromagnetic fields that is represented by 
a vector potential 
${\bf A}(t, {\bf r})=- {\bf E} t + \frac{1}{2}{\bf B}\times{\bf r}$. 
Here, the electric field is weak i.e. $|{\bf E}|<<1$ so that  
the vector potential changes adiabatically. We use the adiabatic 
approximation and we consider the eigenstates of the 
Hamiltonian $H(t)$ at fixed $t$. We adjust 
the flux $\phi$ through the unit cell of 
the crystal becomes rational i.e. $\phi =(p/q)(h c / e)$. 
One can take an enlarged primitive lattice vectors, such as ${\bf e}_x$ and 
$q {\bf e}_y$, and then, the system has symmetric under 
the magnetic translation in terms of 
an enlarged unit cell (the magnetic unit cell) 
$0\leq k_x<2 \pi / a, 0\leq k_y <2 \pi / q a$. The eigenfunctions
of the system $\Phi_{\bf k}(t,{\bf r})$ is in the Bloch state 
$\Phi_{\bf k}(t,{\bf r})=e^{i {\bf k}\cdot{\bf r}}u_{\bf k}({\bf r})$, 
where ${\bf k}$ is on the magnetic Brillouin Zone (MBZ) 
and the function $u_{\bf k}(t, {\bf r})$ obey the generalized 
Bloch conditions. The band splitting occurs 
and one obtains $q$ magnetic subbands. 
The ``Hamiltonian'' for 
$u_{\bf k}(t, {\bf r})$ is written $H_{\bf k}(t)$. Because of 
the gauge invariance, one can see that the adiabatic 
parameter $- {\bf E} t$ is on the MBZ. i.e. one may write 
$H_{\bf k}(t)=H_{{\bf k}-{\bf E}t}$. We should note that 
$H_{\bf k}(t)$ is compactified on the MBZ. So, we can introduce the 
periodicity $T$ for time when ${\bf E}//{\bf G}$, where ${\bf G}$ is the 
reciprocal lattice vector for the enlarged Bravais lattice.   
For example, $T=h/e E a$ for ${\bf E}//{\bf e}_x$ and $T=h / e E q a$ 
for ${\bf E}//{\bf e}_y$. Then, the Berry phase arises 
when one solves the time-dependent Shr\"{o}dinger equation 
by using the adiabatic approximation. We define the electric
polarization ${\bf P}(t)$ and calculate its change per a cycle 
of the adiabatic process, $\Delta{\bf P}=\int_0^{T} dt \dot{\bf P}(t)$.  
One can show that $\dot{P(t)}$ is equivalent to the QH
current when the Fermi level lies in the energy gap, i.e. 
$\dot{\bf P}(t)=\sigma_{xy}{\bf E}\times{\bf e}_z.$ 
The Hall conductance $\sigma_{xy}$ is written by the Chern number 
and is quantized in the multiple of $e^2 / h$. 
It has been pointed out 
that the Chern number is written by the Berry phase, 
and then, $\Delta {\bf P}$ is closely related to the Berry phase. It 
is obtained by the adiabatic process and its direction is perpendicular 
to the electric field. Then, it is quite different with 
the usual induced polarization under an electric field. $\Delta {\bf P}$
is similar to the spontaneous polarization in crystalline
dielectrics which is induced adiabatically change of the Kohm-Sham
potential and written by the Berry phase. 
In the system, Hamiltonian $H_{\bf k}(t)$ changes ac-like, and 
the current is dc. $\Delta {\bf P}$ which corresponds to the 
charge transfer density per cycle does 
not depend on $T$ and quantized. The fact is analogous 
to the adiabatic pumping in 1D. 

The discussion can be extended to the 3D system. 
In 3D, the QHE also occurs in the ``rational'' 
magnetic field.  
Recent argument by Koshino et. al. pointed out that it may be possible 
to realize the 3D QHE in the magnetic field 
around 400 Tesla in the organic compounds (TMTSF)$_2$X. 
As well as 2D, the polarization change per cycle 
is equivalent to the time integral of 3D quantized Hall current, whose 
conductivity is represented by the three sets of the Chern number. 
It was shown that the conductivity, and also 
the polarization change is written by the Berry phase.

\section{Berry phase and spin quantum Hall effect in the vortex 
state of superfluid $^3$He in 2D}
A parallel discussion can be made in the 
vortex state of rotating superfluid $^3$He in 2D. 
Superfluid in rotation is an direct analogy of 
the type II supercondcutors with infinite London penetration depth.  
In 2D, A-phase is stabilized by the boundary effect 
and spin currents are well defined.  
we introduce adiabatically changing vector potential 
that couples to spin current. This coupling is 
equivalent to the Zeeman coupling between spin and 
 a magnetic field with weak and homogeneous gradient in the 
rotating frame. 
The system has periodicity in terms of 
an enlarged unit cell ({\it not} the unit cell of the 
vortex lattice. See, Fig. 1), which is the 
direct analogy of the magnetic unit cell. Then,  
quasiparticles are in the Bloch states.  
\begin{figure}[h]
\begin{center}\leavevmode
\includegraphics[width=0.5\linewidth]{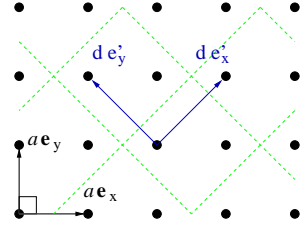}
\caption{ 
The region surrounded by green dotted lines is the enlarged unit cell. 
Black dots shows vortices. 
}\end{center}
\end{figure}
The adiabatic parameter is on the torus as well 
as the arguments in the previous section and the process 
can be closed. Then, a Berry phase is generated. 
When quasiparticles has an excitation gap, 
spin current flows perpendicular to the gradient field 
and quantized, i.e. spin quantum Hall effect (SQHE) 
occurs. The SQHE has been argued by 
Volovik and Yakovenko in $^3$He-A without rotation, and 
Vishwanath, and Vafek et.al. in the vortex state of $d$-wave 
superconductors. 
The authors of the present paper have shown that 
the conductivity in the vortex state of superfluid $^3$He 
is written by the Chern number as well as $d$-wave case,  
and moreover, closely related to the Berry phase. 
In the system, Hamiltonian changes ac-like and spin current is dc. 
The spin transfer $\Delta S_z$ per 
a cycle of the process per a unit cell is quantized 
and does not depends on the period of the cycle. Therefore, 
the effect is analogous to the adiabatic pumping. 
The magnetization change is defined as $\Delta S_z / d$ 
($d$: the length of the boundary of the enlarged unit cell. See, Fig. 1).  
Obviously, it is written by the Berry phase and
analogous to the spontaneous polarization in the 
crystalline dielectrics.    
\begin{ack}
The authors thank H. Aoki, T. Aoyama, K. Ishikawa, 
N. Maeda, K. Maki, M. Sato, Z. Te\v{s}anovi\'{c} and 
F. Zhou for useful discussions. 
\end{ack}
%
%

\end{document}